\documentclass[10pt,aps,prb,twocolumn,groupedaddress]{revtex4-1}
\usepackage{physics}
\usepackage{amsmath, amssymb}
\usepackage[english]{babel}
\usepackage{graphicx}
\usepackage{url}

\newcommand{\Lagr}{\mathcal{L}}

\begin{document}

%\preprint{}

\title{Block product density matrix embedding theory for strongly correlated spin systems}


\author{Klaas Gunst}
\email[]{Klaas.Gunst@UGent.be}
%\thanks{}
%\affiliation{}

\author{Sebastian Wouters}
%\affiliation{}

\author{Stijn De Baerdemacker}
%\affiliation{}

\author{Dimitri Van Neck}
\affiliation{Center for Molecular Modeling, Ghent University, Technologiepark 903, 9052 Zwijnaarde, Belgium}


\date{\today}

\begin{abstract}
Density matrix embedding theory (DMET) is a relatively new technique for the calculation of strongly correlated systems. Recently block product DMET (BPDMET) was introduced for the study of spin systems such as the anti-ferromagnetic $J_1-J_2$ model on the square lattice. In this paper, we extend the variational ansatz of BPDMET using spin-state optimization, yielding improved results. We apply the same techniques to the Kitaev-Heisenberg model on the honeycomb lattice, comparing the results when using several types of clusters. Energy profiles and correlation functions are investigated. A diagonalization in the tangent space of the variational approach yields information on the excited states and the corresponding spectral functions. 
\end{abstract}

%\keywords{}

\maketitle


\section{\label{sec:Introduction}Introduction}
When studying quantum many-body systems, exactly solving the system becomes unfeasible for large system sizes. The exponential scaling of the Hilbert space dimension with the number of particles inhibits an exact simulation of larger systems, and approximate methods have to be used.

A particular type of quantum many-body systems is the quantum spin-lattice system. For this system, interacting spins are localized on the lattice points of a lattice. In this paper, we extend the block product matrix embedding theory (BPDMET),\cite{Fan-2015} a method that was recently introduced to study quantum spin-lattices. We study the validity of the model by applying it to the $J_1-J_2$ model on the square lattice with Heisenberg interaction and the Kitaev-Heisenberg model. These are two spin lattice systems of particular interest.

The first system has been a long time subject of research. This partly because of its fundamental interest in its simplicity, but also for its use in Fe-based superconductors and other materials. For example, high-$T_c$ superconductivity in iron pnictide (or oxypnictides) has been discovered with LaOFeAs being the first.\cite{Si-2008} The Fe atoms form a square lattice in these iron pnictides and they exhibit NN and NNN superexchange interactions that can be described with this $J_1-J_2$ model (however with $S = 1$ or $2$). When the crystal is doped, an effective model with a $t-J_1-J_2$ Hamiltonian can be suggested, introducing a kinetic component.\cite{Si-2008} This should give rise to superconductivity. Also the properties of $\mathrm{Li}_2\mathrm{VOSiO}_4$ have been investigated through the $J_1-J_2$ model.\cite{Melzi-2001} 

The Kitaev-Heisenberg model was first introduced for the theoretical examination of iridium oxides of the form $A_2\mathrm{IrO_3}$, with $A = \mathrm{Li,\,Na}$.\cite{Chaloupka-2010} Herein, $\mathrm{Ir^{4+}}$ ions form honeycomb-like lattice planes and have an effective spin one-half. The interaction between the different effective spins is anisotropic. Experimental evidence has shown that the proposed Kitaev-Heisenberg model is a successful model for the iridium oxides, however some extensions of the model have been introduced for a better description.\cite{Kimchi-2011, Singh-2012, Chaloupka-2013, Reuther-2014, Rau-2014, Chaloupka-2015} Particular interest for the Kitaev-Heisenberg model has arisen since the model is able to have a Kitaev spin liquid\cite{Kitaev-2006} as ground state within a finite parameter region.

To solve these systems, exact diagonalization is unfeasible for larger systems and approximate methods have to be used. Some examples of approximate methods are series expansion,\cite{Gelfand-1989, Singh-1999, Kotov-1999, Oitmaa-2015} large-N expansion,\cite{Read-1991} density matrix renormalization group,\cite{White-2007, Yan-2011, Jiang-2012} projected entangled pair states,\cite{Murg-2009} and coupled cluster methods.\cite{Darradi-2008}

An approximate solution of quantum many body systems can also be obtained through embedding theories. 
%Suppose one is only interested in a small part of the whole system. This part can be a specific region in space, but any other division of the total Hilbert space of the system into subspaces is also possible. By dividing the total system into two parts,
Here, the system is divided into two parts: an \textit{impurity}, \textit{cluster} or \textit{fragment} (which is the subsystem of interest) and an \textit{environment}. Using this division, embedding theories are able to solve the problem approximately.\cite{Sun-2016}

The total Hilbert space of the system is now a direct product of the Hilbert spaces of the impurity and the environment (with dimension $A$ and $B$ respectively). A basis for this Hilbert space is given by $\{\ket{\alpha_i} \otimes \ket{\beta_j}\}$, where $\ket{\alpha_i}$ are states of the impurity and $\ket{\beta_j}$ are states restricted to the environment. Every state $\ket{\Psi}$ in the total Hilbert space can be written as 

\begin{equation}
\ket{\Psi} = \sum\limits_{i=1}^A \sum\limits_{j=1}^B \Psi_{ij} \ket{\alpha_i}\ket{\beta_j}
=\sum\limits_k^{\mathrm{min}(A,B)} \lambda_k \ket{\widetilde{\alpha}_k}\ket*{\widetilde{\beta}_k}.
\label{eq:Schmidt_intro}
\end{equation}
The latter result is obtained by using a singular value decomposition of the matrix $\Psi_{ij}$.
This is known as the Schmidt decomposition of a state.
Here, $\ket{\widetilde{\alpha}_k}$ are states of the impurity and $\ket*{\widetilde{\beta}_k}$ are states of the environment. The summation is restricted to the minimum of the impurity and the environment dimension. Since the dimension of the environment is typically much larger than the dimension of the impurity, the summation is limited by the impurity. It is thus clear that only $A$ states in the environment are needed for the construction of the wave function. If only one of the singular values $\lambda_k$ is nonzero, the state $\ket{\Psi}$ can be factorized and impurity and environment are not entangled. However, if several singular values are nonzero, $\ket{\Psi}$ is called entangled.\cite{Bennett-1996, Horodecki-2009, Eisert-2010, Eisert-2013}

Embedding theories capitalize on this division of the system. By replacing the environment by an approximate model, one tries to calculate the properties of the impurity accurately and cost-effectively. The simplest option is to approximate the environment in such a way that there is no entanglement with the impurity. This is a good approximation when the Schmidt singular values have one dominant non-zero value. For many systems this is sufficient. However, for systems with strong static correlation between the impurity and environment one has to go beyond the mean-field approximation. Static correlation refers to systems where a product state is not sufficient for a qualitative description of the system, but a superposition of multiple product states is needed, i.e. whenever substantial entanglement is present between the impurity and environment, implying several important Schmidt values $\lambda_k$ in Eq.~(\ref{eq:Schmidt_intro}).

One of the more powerful and popular embedding theories is dynamical mean-field theory (DMFT)\cite{Metzner-1989,Georges-1992,Georges-1996,Zgid-2011}. It maps the system to an impurity and a non-interacting bath in a self-consistent way, using the single-particle Green's function.

Density Matrix Embedding Theory (DMET) is a new embedding theory first proposed by Knizia and Chan\cite{Knizia-2012} for the Hubbard model. It was later extended to full quantum chemical Hamiltonians.\cite{Knizia-2013} For ground state energies, DMET is a computationally cheaper alternative to DMFT with similar accuracy. The self-consistency for DMET is based on the density matrix, instead of the frequency dependent Green's function in DMFT. Information about excited states in DMET can still be explored.\cite{Booth-2015, Wouters-5yr}

In DMET the entanglement between impurity and environment is explicitly kept and the wave function is of the form given by Eq.~(\ref{eq:Schmidt_intro}). Finding the Schmidt basis for the environment $\{\ket*{\widetilde{\beta}_k}\}$ can be done if the exact wave function $\ket{\Psi}$ is known. This is, however, not an option since finding the exact wave function is equivalent to solving the many-body problem. DMET solves the lack of a priori knowledge of $\ket{\Psi}$ by embedding the impurity in an approximate bath. Solving this combined impurity and bath system is called the embedded problem. To find this bath space, one can use different techniques. A Fock space of bath orbitals which is  obtained from a low-level particle-number conserving mean-field wave function is used in the original Refs. \onlinecite{Knizia-2012, Knizia-2013}  and is illustrated extensively in Ref. \onlinecite{Wouters-5yr}. Other methods are also possible: single-particle states from Hartree-Fock-Bogoliubov theory\cite{Zheng-2016,LeBlanc-2015} and antisymmetrized geminal power (AGP) wave functions\cite{Tsuchimochi-2015} have also been used. Extensions of DMET to coupled interacting fermion-boson systems through coherent state wave functions for phonons have been described as well.\cite{Sandhoefer-2016}

Adapting DMET for spin lattices has given rise to the so-called cluster density matrix embedding theory (CDMET), as introduced by Fan \textit{et al.}\cite{Fan-2015} In this paper, however, we opt for the name block product DMET (BPDMET) in order to avoid possible confusion with fermionic DMET when the impurity consists of a cluster of degrees of freedom. In BPDMET, bath states in a spin lattice system are represented by block product states, which is emphasised by our alternative name. Recently, this method has been further extended by implementing BPDMET with the hierarchical mean-field approach.\cite{Qin-2016} In the present work, the original BPDMET is used and the ansatz is further extended with so-called spin-state superpositions in the impurity, yielding improved results.

As will be shown, the BPDMET ansatz\cite{Fan-2015} can easily be written as a particular case of a more general tensor network state (TNS).
The concept of tensor network states is an increasingly important technique for the description of highly correlated systems. It can be viewed as an extension of the density matrix renormalization group (DMRG) as introduced by White.\cite{White-1992, White-1993} DMRG was shown to be highly accurate and useful for one-dimensional lattice systems, but also reasonably small two-dimensional lattices can be studied up to high accuracy.\cite{Stoudenmire-2012} The DMRG algorithm has been later rewritten as an optimization of a matrix product state (MPS) ansatz.\cite{Ostlund-1995,Rommer-1997} %Projected entangled pair states or PEPS is the extension of MPS into two and higher dimension.\cite{Verstraete-2004, Murg-2009}. 

The concept of TNS is quite general and includes projected entangled pair states (PEPS)\cite{Verstraete-2004, Murg-2009} which is the extension of MPS into two and higher dimensions, as well as tree TNS (TTNS)\cite{Murg-2010, Murg-2015, Nakatani-2013, Li-2012} and complete-graph TNS (CGTNS).\cite{Marti-2010, Marti-2011} %Other popular TNS are given by the tree TNS (TTNS)\cite{Murg-2010, Murg-2015, Nakatani-2013, Li-2012} and complete-graph TNS (CGTNS)\cite{Marti-2010, Marti-2011}

In section \ref{sec:Method}, the concept of BPDMET is introduced by means of a variational ansatz. The optimization procedure is explained as well as the calculation of the energy and other properties in the BPDMET framework. We also extend the variational ansatz using spin-state superposition in the impurity. Finally, we point out a possible way of extracting information on spectral properties.  The link with tensor network states is also made. In section \ref{sec:Results}, the method is applied to the 2D Heisenberg model on a square lattice and to the Kitaev-Heisenberg model on a honeycomb lattice. Results for the energies, correlation functions and the location of quantum phase transitions are studied. The concept of diagonalization in tangent space and the resulting spectral function is applied to both lattice systems. Summary and conclusions are provided in section \ref{sec:Conclusions}.

\section{\label{sec:Method}Method}
\subsection{Block Product DMET}

The BPDMET method as introduced by Fan \textit{et al.}\cite{Fan-2015} can be used for the approximate solution of spin lattice systems. These systems can be used to model magnetic properties of materials. A spin lattice system comprises a number of lattice sites $N$. Each site has a spin degree of freedom interacting with other spins. Only spin-spin interactions are investigated in this paper. For these systems, the total Hamiltonian can be written in its most general form as
\begin{equation}
\hat{H} = \sum\limits_{m,n} \sum\limits_{\mu,\nu} J_{\mu \nu}^{mn}\hat{S}^\mu_m\hat{S}^\nu_n.
\label{eq:generalHam}
\end{equation}
Here, $m$ and $n$ denote lattice indices and $\mu$ and $\nu$ denote the spatial components $x$, $y$ and $z$, i.e. the different measuring directions for the spin operator.
Magnetic terms have been excluded but can be introduced in a straightforward manner.
 
\begin{figure}[ht]
\includegraphics[width = 0.3\textwidth]{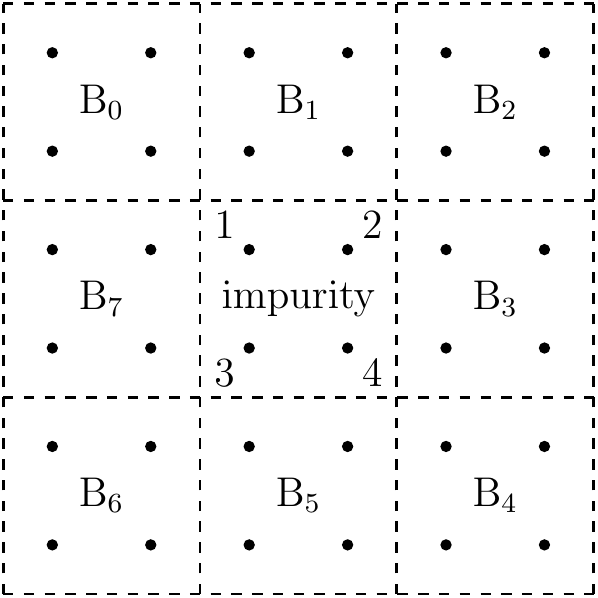}
\caption{An exemplary division of a square lattice into 1 impurity and 8 bath clusters of size $2\times 2$.}  
\label{fig:bath_division}
\end{figure}

An interesting variational ansatz for the system was provided by Fan \textit{et al.}\cite{Fan-2015} After division of the spin lattice into an impurity and environment, they propose a replacement of the exact environmental states $\ket*{\widetilde{\beta}_k}$ in Eq.~(\ref{eq:Schmidt_intro}) by a set of block product states $\ket{\mathrm{BPS}_k}$. With this approximation, the wave function of the impurity model becomes:
\begin{equation}
\ket{\Psi} = \sum\limits_i a_i \ket{\alpha_i}\ket{\mathrm{BPS}_i},
\label{eq:Schmidtapprox}
\end{equation}
where $i$ labels the different states in the Hilbert space of the impurity. To define these block product states, the spin lattice system is divided into different  clusters. One of the clusters is the impurity. The other clusters are called the bath clusters. An exemplary division can be seen in Fig.~\ref{fig:bath_division}. With this division of the lattice system, the block product states are defined as follows:
\begin{equation}
\ket{\mathrm{BPS}_i} = \prod\limits_{C \in \mathrm{bath\, clusters}} \sum\limits_\beta b^i_{C\beta}\ket{\beta}_C.
\label{eq:BPS}
\end{equation}
Here $\{\ket{\beta}_C\}$ is a complete set of states within the Hilbert space restricted to the bath cluster $C$. For example, if each bath cluster contains 3 sites with spin-$\frac{1}{2}$, we have $\{\ket{\downarrow\downarrow\downarrow},\ket{\downarrow\downarrow\uparrow},\ket{\downarrow\uparrow\downarrow},..,\ket{\uparrow\uparrow\uparrow}\}$ in the natural basis (eigenstates of $\hat{S}^z$). These block-product states have to be optimized so that Eq.~(\ref{eq:Schmidtapprox}) optimally represents the exact ground state wave function. The approximation made is that the correlations \textit{within} the bath clusters are fully taken into account, while the correlations \textit{between} the bath clusters are only taken into account via mediation of the impurity.

The dimension of the complete Hilbert space is given by $2^N$ with $N$ the number of spins, and hence grows exponentially with the number of spins. When the size of all clusters is chosen equal, the number of degrees of freedom of the BPDMET ansatz is given by $2^{N_s}(2^{N_s} -1)(N_C - 1) + 2^{N_s}$. $N_s$ is the number of spins in a cluster and $N_C$ is the number of clusters. When the cluster size $N_s$ is kept fixed, the degrees of freedom scale linearly with the number of spins (or the number of clusters $N_C$). This linear scaling clearly is the major advantage of the BPDMET ansatz.

\subsection{\label{sec:opt}Optimizing the wave function}

Thanks to the linear scaling, one can target the state $\ket{\Psi}$ (Eq.~(\ref{eq:BPS})) variationally. Optimization of the block product states and finding the ground state of the impurity model proceeds in an iterative way. At each step of the iteration, a bath cluster is chosen and its state corresponding to a certain impurity state is optimized. This way, a large number of coefficients of the variational wave function is kept fixed, and only a restricted number of coefficients is optimized in each step. The variational wave function within the impurity model can be written as
\begin{equation}
\ket{\Psi} = \sum\limits_i a_i \ket{\alpha_i} \left(\prod\limits_{C\in\mathrm{bath\, cl.}} \sum\limits_\beta b^i_{C\beta} \ket{\beta}_C\right),
\label{eq:var_wave}
\end{equation}
with $C$ being the different bath clusters. For every possible wave function of this form, we can take
\begin{equation}
\sum\limits_\beta b^{i*}_{C\beta}b^i_{C\beta}=1,
\label{eq:norm_b}
\end{equation} 
by absorbing appropriate factors in the $a_i$'s.
Even more, when the wave function is normalized, 
\begin{equation}
\sum\limits_i a_i^*a_i = 1
\label{eq:norm_a}
\end{equation}
will also be satisfied.

The coefficients of the variational wave function are obtained with a restricted optimization. All coefficients are fixed, except for the $a_i$'s and $b^{i_0}_{B_0 \beta}$, i.e. the $b$-coefficients corresponding with the impurity state $i_0$ and a chosen bath cluster $B_0$. By looping over the different $i_0$'s and bath clusters $B_0$ we optimize the DMET wave function iteratively.

We now rewrite the wave function given by Eq. (\ref{eq:var_wave}) as
\begin{equation}
\begin{split}
\ket{\Psi} = &\sum\limits_{i (\neq i_0)} a_i \ket{\alpha_i} \prod\limits_C \sum\limits_\beta b^i_{C\beta} \ket{\beta}_C + \\
&\sum\limits_{\beta}  a_{i_0} b^{i_0}_{B_0\beta}\ket{\alpha_{i_0}}\ket{\beta}_{B_0}\prod\limits_{C (\neq B_0)} \sum\limits_{\beta'} b^{i_0}_{C\beta'}\ket{\beta'}_C.
\end{split}
\label{eq:basis}
\end{equation}
Since optimization happens over $a_i$ with $i \neq i_0$ and $a_{i_0}b^{i_0}_{B_0\beta'}$, every iteration is equivalent to a diagonalization in the low-dimensional subspace spanned by $\{\ket{\phi_\alpha},\ket{\phi_\beta}\}$, with
\begin{align*}
\ket{\phi_{\alpha_i}} &= \ket{\alpha_i} \prod\limits_C \sum\limits_\beta b^i_{C\beta} \ket{\beta}_C,\\
\ket{\phi_\beta} &= \ket{\alpha_{i_0}}\ket{\beta}_{B_0}\prod\limits_{C (\neq B_0)} \sum\limits_{\beta'} b^{i_0}_{C\beta'} \ket{\beta'}_C.
\end{align*}

To find the optimal solution with every iteration, the following Lagrangian is minimized within this restricted Hilbert space:
\begin{equation}
\Lagr = \expval{\hat{H}}{\Psi} - \lambda \braket{\Psi},
\label{eq:Lagrangian}
\end{equation}
yielding a linear eigenvalue problem.

The Lagrangian multiplier $\lambda$ is the variational energy of the wave function. Within every iteration, the solution corresponding to the smallest $\lambda$ is chosen. It is clear that the solution of the previous iteration can still be chosen within the freedom of the parameters in the current iteration. Because of this, the minimal $\lambda$-value obtained in the current iteration is at least as small as the $\lambda$ value of the previous iteration. Since $\lambda$ decreases with every two consecutive iterations, we converge to a minimal $\lambda$ value, although it is not guaranteed to be the global minimum of the energy. The complexity of a major iteration (i.e. an iteration over all $i_0$-values and over all bath clusters $B_0$) is of the order $\mathcal{O}(N_C^2)$ with $N_C$ the number of clusters. The number of major iterations needed up to convergence may increase when increasing the number of clusters. However, the scaling of the problem when enlarging the number of spins is of course more favorable than the exponential scaling of the exact diagonalization, as long as the size of the clusters does not change. 

When changing the size of the clusters, the algorithm scales exponentially due to the exponential scaling of the restricted Hilbert space chosen every minor iteration step. Even more, the number of impurity states also grows exponentially, making the number of minor iteration steps in every major iteration step also blow up. The scaling of every major iteration step is approximately $\mathcal{O}(2^{4N_s})$. Making the clusters bigger results quickly in high computational times.

Details of the calculations are investigated in more depth in the Supplemental Material.\cite{supplemental}

\subsection{\label{sec:sso}Spin-state superposition in the impurity}

When looking at the Schmidt decomposition of the exact wave function in a small impurity and a larger environment (Eq.~(\ref{eq:Schmidt_intro})), the set of impurity states $\{\ket{\alpha_i}\}$ can be any basis of the Hilbert space restricted to the impurity. Corresponding environmental states will always be found. When approximating the environmental states by block product states, this freedom of choice is lost. The choice of the impurity states influences the corner of the Hilbert space the BPDMET algorithm optimizes in.
In the original BPDMET method,\cite{Fan-2015} the states of the impurity $\{\ket{\alpha_i}\}$ are given by the natural basis for a spin system, e.g. $\{\ket{\downarrow\cdots\downarrow\downarrow},\ket{\downarrow\cdots\downarrow\uparrow},\ket{\downarrow\cdots\uparrow\downarrow},\cdots,\ket{\uparrow\cdots\uparrow\uparrow}\}$ for $S = \frac{1}{2}$. This choice is arbitrary and influences the obtained results. In this section we extend the BPDMET wave function enabling it to find an optimal set of orthonormal impurity states. Orthonormality is imposed in order to keep the simplifications in the calculation of $\mel{\frac{\partial \Psi}{\partial z*}}{\hat{H}}{\Psi}$, as described in the Supplemental Material.\cite{supplemental}

The adapted BPDMET ansatz is again given by Eq.~(\ref{eq:Schmidtapprox}). However, the impurity states are now superpositions of the natural basis states: $\ket{\alpha_i} = \sum\limits_m U_{im} \ket{m}$ with $\ket{m} = \{\ket{\downarrow\cdots\downarrow\downarrow},\ket{\downarrow\cdots\downarrow\uparrow},\cdots\}$ and $U$ a unitary matrix. 
The BPDMET ansatz is thus given by:
\begin{equation}
\ket{\Psi} = \sum\limits_i a_i \sum\limits_m U_{im} \ket{m}\prod\limits_{C \in \mathrm{bath\, cl.}} \sum\limits_\beta b^i_{C\beta}\ket{\beta}_C.
\label{eq:BPDMETfull}
\end{equation}
Now the unitary matrix has to be optimized as well. We simply extend the original algorithm:  the block-product states and the unitary matrix are optimized successively until convergence is obtained.

The unitary matrix is optimized by minimizing the variational energy $\lambda$ through successive Jacobi-rotations.\cite{Poelmans-2015}

The optimization scheme for BPDMET with spin-state superposition now looks like:
\begin{enumerate}
\item Initialization of of the impurity states $(U)$ and the BPS $(b^i_{C\beta})$
\item Optimization of the BPDMET wave function:
	\begin{enumerate}
	\item \label{itm:BPS} Optimization of the BPS: Loop over bath clusters $B_0$ and impurity states $i_0$ and solve Eq.~(\ref{eq:Lagrangian}) keeping appropriate parameters fixed until convergence
	\item \label{itm:spinstate} Optimization of the impurity states through successive Jacobi rotations.
		\item Restart from step \ref{itm:BPS} until convergence.
	\end{enumerate}
\end{enumerate}
As a convergence criterion both the variational energy $\lambda$ and the BPDMET energy $E$ can be used. The BPDMET energy (Eq.~\ref{eq:DMETenergy}) is an alternative way to calculate the energy of the system within the DMET framework and will be introduced in the next section. In this paper we choose the BPDMET energy, as it converges somewhat more slowly than the variational energy. The faster convergence of the variational energy is clear since it is quadratically dependent on the error of the wave function, while generally, the expectation value of a property (which Eq.~(\ref{eq:DMETenergy}) represents) has a linear dependency as shown in the appendix.

\subsection{\label{sec:expval}Expectation values}

In this section, the calculation of expectation values within the BPDMET framework is discussed. The method of calculation is equivalent to the method used in DMET as presented by Wouters \textit{et al}.\cite{Wouters-2016} In BPDMET, we divide the lattice into different clusters and choose one cluster as the impurity cluster. In this paper the division happens in such way that all clusters are equivalent with respect to the lattice symmetry. All clusters can be transformed into each other by using a translation or rotation for which the lattice is invariant. By picking one cluster $C$ as impurity and calculating its corresponding BPDMET wave function $\ket{\Psi_C}$, we immediately know all the BPDMET wave functions corresponding to the other choices of the impurity. This is a great advantage as far as computational time is concerned.

Wouters \textit{et al.}\cite{Wouters-2016} noted the fundamental difference between \textit{local} and \textit{nonlocal} operators. Local operators act within one impurity while nonlocal operators do not.
Just like in the original DMET framework, expectation values for local operators are quite straightforward, while expectation values for nonlocal operators require some inventiveness. When a local operator $\hat{A}$ only acts upon cluster $C$, its expectation value can be calculated by:
$\ev*{\hat{A}} = \ev{\hat{A}}{\Psi_C}$, where $\ket{\Psi_C}$ is the calculated BPDMET wave function with cluster $C$ chosen as impurity.
Note however that also operators consisting of summations of local operators impose no problem. For example, the expectation value of the total spin in the $z$-direction is given by:
\begin{equation}
\ev{\hat{S}_\mathrm{tot}^z} = \sum\limits_C \ev{\hat{S}_\mathrm{tot_C}^z}{\Psi_C} = N_C \ev{\hat{S}_\mathrm{tot_C}^z}{\Psi_C},
\end{equation}
where $\hat{S}_\mathrm{tot_C}^z$ is the total spin in the $z$-direction restricted to sites belonging to cluster $C$. Since all $\ket{\Psi_C}$ are equivalent, the summation over the different clusters is omitted in the last step, where $N_C$ is the number of clusters in the system.

For nonlocal operators the original DMET framework suggests splitting these operators appropriately.\cite{Wouters-2016} The expectation values of interest for the spin lattice systems studied in this paper are given by a summation of scalar products of spin operators. The expectation value can thus be written as the sum of the expectation values of the different terms $\ev{\hat{\mathbf{S}}_i \cdot \hat{\mathbf{S}}_j}$. When both $i$ and $j$ are sites within one cluster, this expectation value is an expectation value of a local operator. However, when this is not the case, this expectation value is an expectation value of a nonlocal operator and will be calculated as:
\begin{equation}
\ev{\hat{\mathbf{S}}_i \cdot \hat{\mathbf{S}}_j} = \frac{1}{2}\ev*{\hat{\mathbf{S}}_j \cdot \hat{\mathbf{S}}_i}{\Psi_{C_i}} + \frac{1}{2}\ev*{\hat{\mathbf{S}}_i \cdot \hat{\mathbf{S}}_j}{\Psi_{C_j}},
\end{equation}
where $\ket{\Psi_{C_i}}$ and $\ket*{\Psi_{C_j}}$ are the BPDMET solutions with the impurity chosen to be the cluster of site $i$ or site $j$, respectively. Since the solutions of the BPDMET for different impurity clusters are equivalent, the calculation of expectation values of operators that respect the lattice symmetry can be simplified. Examples of these are the Hamiltonian $\hat{H}$ in Eq.~(\ref{eq:generalHam}) and the squared total spin $\hat{\mathbf{S}}^2_\mathrm{tot}$. These are given by:
\begin{align}
\ev{\hat{H}} &= N_C \sum\limits_{i \in \mathrm{C,\,}j} J_{ij} \ev**{ \hat{\mathbf{S}}_i \cdot \hat{\mathbf{S}}_j}{\Psi_C},
\label{eq:DMETenergy}
\\
\ev{\hat{\mathbf{S}}^2_{\mathrm{tot}}} &= N_C \sum\limits_{i \in \mathrm{C,\,}j}\ev**{ \hat{\mathbf{S}}_i \cdot \hat{\mathbf{S}}_j}{\Psi_C}.
\end{align}

By calculating properties in this way, we take into account that BPDMET describes the impurity more accurately than the bath. However, this method is not variational in nature, so energies obtained through Eq.~(\ref{eq:DMETenergy}) are not upper bounds to the exact energy, and squared total spins can be slightly negative. From now on, this non-variational energy will be called the BPDMET energy and denoted by $E$, while the variational energy is given by the Lagrangian multiplier $\lambda$ in Eq.~(\ref{eq:Lagrangian}).

\subsection{\label{sec:tangent}Tangent space and excitations}

The BPDMET algorithm can also be extended for the calculation of approximate spectral functions. 
For the calculation of the spectral function, the Hamiltonian restricted to the tangent space of the BPDMET ansatz (Eq. \ref{eq:BPDMETfull}) is diagonalized. Shifting of weights between the different $b^i_{C\beta}$ and $U_{im}$ parameters is possible without changing the actual wave function. It is thus clear that the parametrization is redundant. By introducing a set of restrictions, the redundancy can be lifted and normalization of the wave function can be imposed:
\begin{align}
\sum\limits_\beta b^{i*}_{C\beta}b^i_{C\beta}&=1 \qquad \forall i, C\\
\sum\limits_j U^*_{ij} U_{ij} &=1 \qquad \forall i\\
\sum\limits_i a_i^* a_i &=1.
\end{align}
The tangent space is  constructed by taking these restrictions into account and differentiating with respect to the nonredundant parameters. Diagonalization of the Hamiltonian restricted to the tangent space now amounts to solving a generalized eigenvalue problem. Eigenvectors with eigenvalues close to zero of the overlap matrix are projected out.

The spectral function is given by
\begin{equation}
A(\omega, \hat{X}) = -\frac{1}{\pi}\Im[\ev{\hat{X}^\dagger \frac{1}{\omega - (\hat{H} - E_0) + i\eta} \hat{X}}{\phi_0}],
\end{equation}
where $\hat{X}$ is a perturbation operator connecting the ground state with the excited states. By restricting to the tangent space, this can be rewritten as
\begin{equation}
A(\omega, \hat{X}) \approx -\frac{1}{\pi}\Im\left[\sum\limits_n \frac{|\mel{\phi_n}{\hat{X}}{\phi_0}|^2}{\omega - (E_n - E_0) + i\eta}\right]
\label{eq:spectral}
\end{equation}
where $\phi_n$ and $E_n$ are the eigenvectors of the generalized eigenvalues problem and their corresponding energies. For $E_n$ and $E_0$, both the variational energy $\lambda$ and the BPDMET energy $E$ in Eq.~(\ref{eq:DMETenergy}) can be used. It will be shown that the latter choice yields inferior results.

In this paper, the spectral function is calculated by searching all the excitations within the tangent space. Another option  is through solving the linear response equation given by:
\begin{equation}
(\omega - (\hat{H} -E_0) + i\eta)\ket{\phi_1} = \hat{X}\ket{\phi_0}.
\end{equation}
Both methods can be done through sparse iterative solvers. However, since the BPDMET ansatz has a rather low number of parameters, the tangent space has a low dimension, making explicit solving feasible.

\subsection{Connection with Tensor Networks}

\begin{figure}[ht]
\includegraphics[width = 0.3\textwidth]{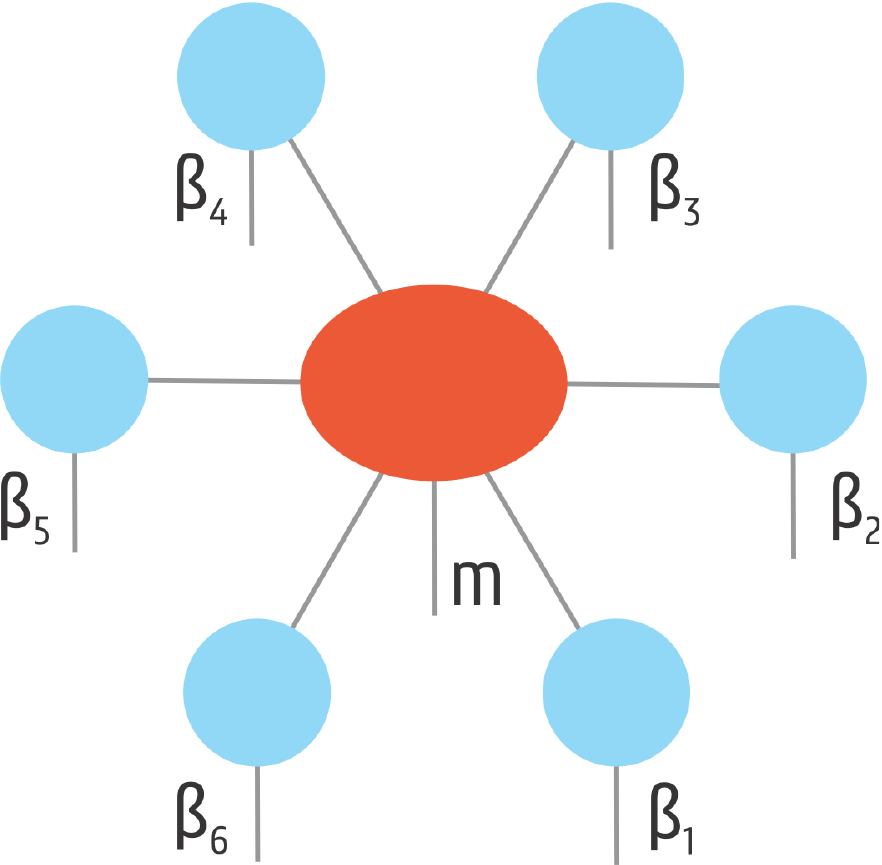}
\caption{(Color online) TNS depiction of the BPDMET ansatz given in Eq.~(\ref{eq:BPDMETfull}). The TNS is given for 6 bath cluster tensors (blue) and one impurity tensor (red). The impurity tensor is given by Eq.~(\ref{eq:imptensor}). The physical indexes are given by the unconnected bonds $(m, \beta_1, \beta_2, \cdots)$.}  
\label{fig:TNS}
\end{figure}

When looking at the BPDMET ansatz, it is clear that it can be represented by the tensor network depicted in Fig.~\ref{fig:TNS}. The central tensor corresponds to the impurity cluster and the remaining tensors to the bath clusters. It should be noted, that these bath tensors can be chosen differently from each other, as can be readily seen from Eq.~(\ref{eq:BPDMETfull}). The spin degrees of freedom are combined into one physical index per cluster. The impurity tensor has virtual indexes connected to every bath tensor. It is therefor a very high rank tensor (one virtual index per bath cluster). However, the impurity tensor is heavily restricted, making the tensor network manageable. The BPDMET high rank impurity tensor can be represented as
\begin{equation}
A^m_{i_1 i_2 i_3 i_4 \cdots} = U_{i_1 m} \delta_{i_1 i_2} \delta_{i_1 i_3} \delta_{i_1 i_4} \cdots
\label{eq:imptensor}
\end{equation}
with $U$ a unitary matrix. The traditional technique to make TNS manageable is the truncation of the virtual dimension (e.g. as in DMRG and PEPS). In BPDMET, the TNS is made manageable by imposing restrictions on the impurity tensor. There are no truncations in the virtual dimension. Clarifying the link with tensor networks can facilitate the  construction of different ansatzes for the bath states.

\section{\label{sec:Results}Results}

In this section, results of the BPDMET method are discussed for two different models. In section \ref{sec:square}, the 2D Heisenberg model on the square lattice with nearest-neighbor (NN) and next-nearest-neighbor (NNN) interaction is studied. The BPDMET results obtained by Fan \textit{et al.} \cite{Fan-2015} are reproduced and compared with the results when spin-state superposition is introduced (see section \ref{sec:sso}). In addition, different order parameters are calculated. In section \ref{sec:kitheis}, the BPDMET algorithm is applied to the Kitaev-Heisenberg model on a Honeycomb lattice\cite{Chaloupka-2010} with 24 spins and compared with exact results.

Spectral functions obtained through diagonalization in the tangent space as discussed in section \ref{sec:tangent} are also presented.

Note that when dividing the complete lattice into two clusters, one impurity and one bath cluster, that the BPDMET ansatz (Eq.~(\ref{eq:BPDMETfull})) yields a redundant parametrization of the complete Hilbert space. In this situation, BPDMET should coincide with exact diagonalization (ED). Comparison with ED for small systems with only one bath cluster allows us to check the correctness of the algorithms and implementations.

\subsection{\label{sec:square}NN and NNN interaction on the square lattice}

The first system under consideration is the square lattice with NN and NNN Heisenberg interactions. The Hamiltonian of this system is given by
\begin{equation}
\hat{H} = J_1 \sum\limits_{\langle i,j \rangle}\hat{S}_i\cdot\hat{S}_j + J_2\sum\limits_{\langle \langle i,k \rangle \rangle}\hat{S}_i\cdot\hat{S}_k
\end{equation}
with  $\langle i,j \rangle$ denoting NN sites, $J_1$ the NN interaction strength, $\langle \langle i,k \rangle \rangle$ denoting NNN sites and $J_2$ the NNN interaction strength. We restrict ourselves to anti ferromagnetic (AF) interactions ($J_1, J_2 \geqslant 0$). This is the same system as investigated by Fan \textit{et al}.\cite{Fan-2015} Three distinct phases are present in the infinite lattice. At low NNN interaction, the ground state is in a N\'eel phase, this is a long-range-ordered phase. At $J_2/J_1 \approx 0.4$ a phase transition from this N\'eel phase happens to a disordered quantum paramagnetic phase. When tuning the system to stronger NNN interactions, the system undergoes a transition to another long range ordered phase at $J_2/J_1 \approx 0.6$. This is the collinear phase. The nature of the intermediate paramagnetic phase is still undecided. Multiple interpretations for this phase have been proposed, such as spin liquids and valence bond states like the columnar and staggered dimer valence bond crystals and the plaquette resonating valence bond (PRVB) state.\cite{Chandra-1988, Gelfand-1989, Read-1989, Takano-2003, Darradi-2008, Murg-2009, Jiang-2012,  Mezzacapo-2012, Yu-2012, LiT-2012, Wang-2013, Hu-2013, Doretto-2014, Gong-2014} In Ref. \onlinecite{Fan-2015}, it is shown that BPDMET calculations suggests no rotational symmetry breaking at the intermediate phase, and evidence is found in favor of the PRVB.

\begin{figure}
\includegraphics[width = 0.47\textwidth]{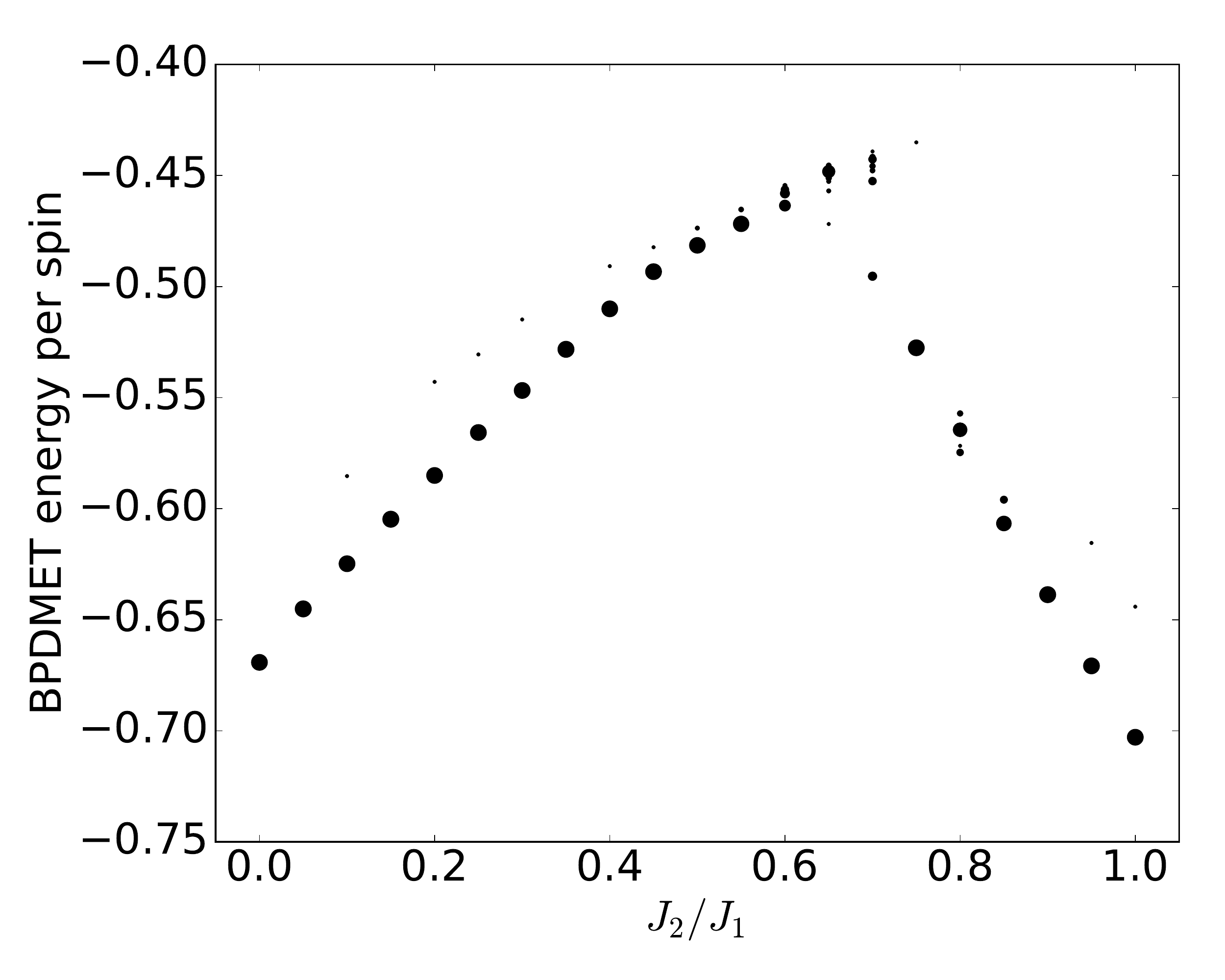}
\caption{Calculations executed with random initialization. At each $J_2/J_1$ value, 40 random calculations are executed. The size of the markers scale with the number of calculations that yielded the corresponding value.}
\label{fig:sq_random}
\end{figure}

\begin{figure}
\includegraphics[width = 0.48\textwidth]{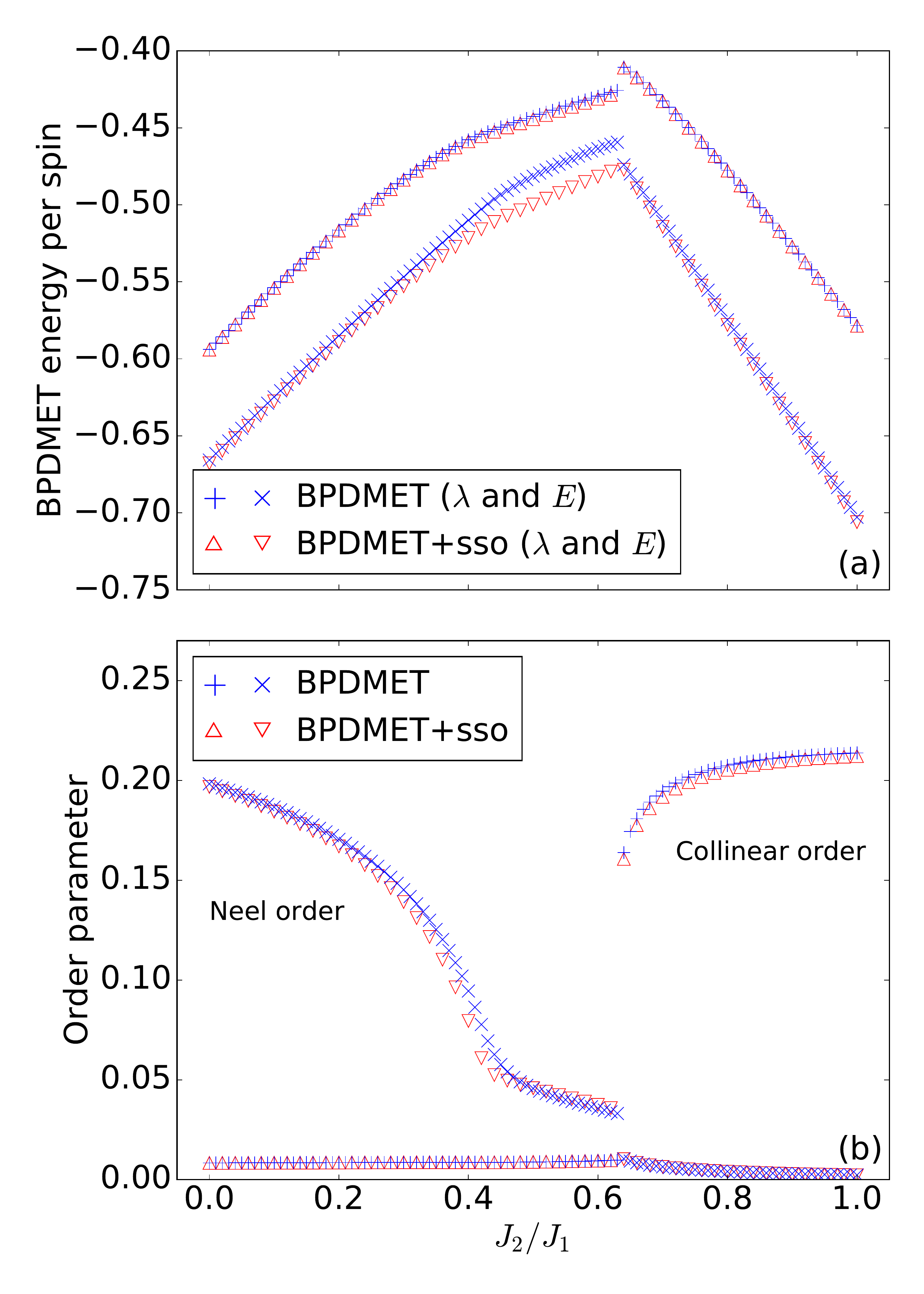}
\caption{\label{fig:sq_64}(Color online) BPDMET calculations with and without spin-state optimization of the $8\times8$ square lattice with NN and NNN interaction. (a) shows the variational $\lambda$ energy ($+$ and $\triangle$) and the non variational BPDMET energy ($\times$ and $\triangledown$). (b) shows the N\'eel ($\times$ and $\triangledown$) and collinear order parameters ($+$ and $\triangle$).}
\end{figure}

\begin{figure}
\includegraphics[width = 0.48\textwidth]{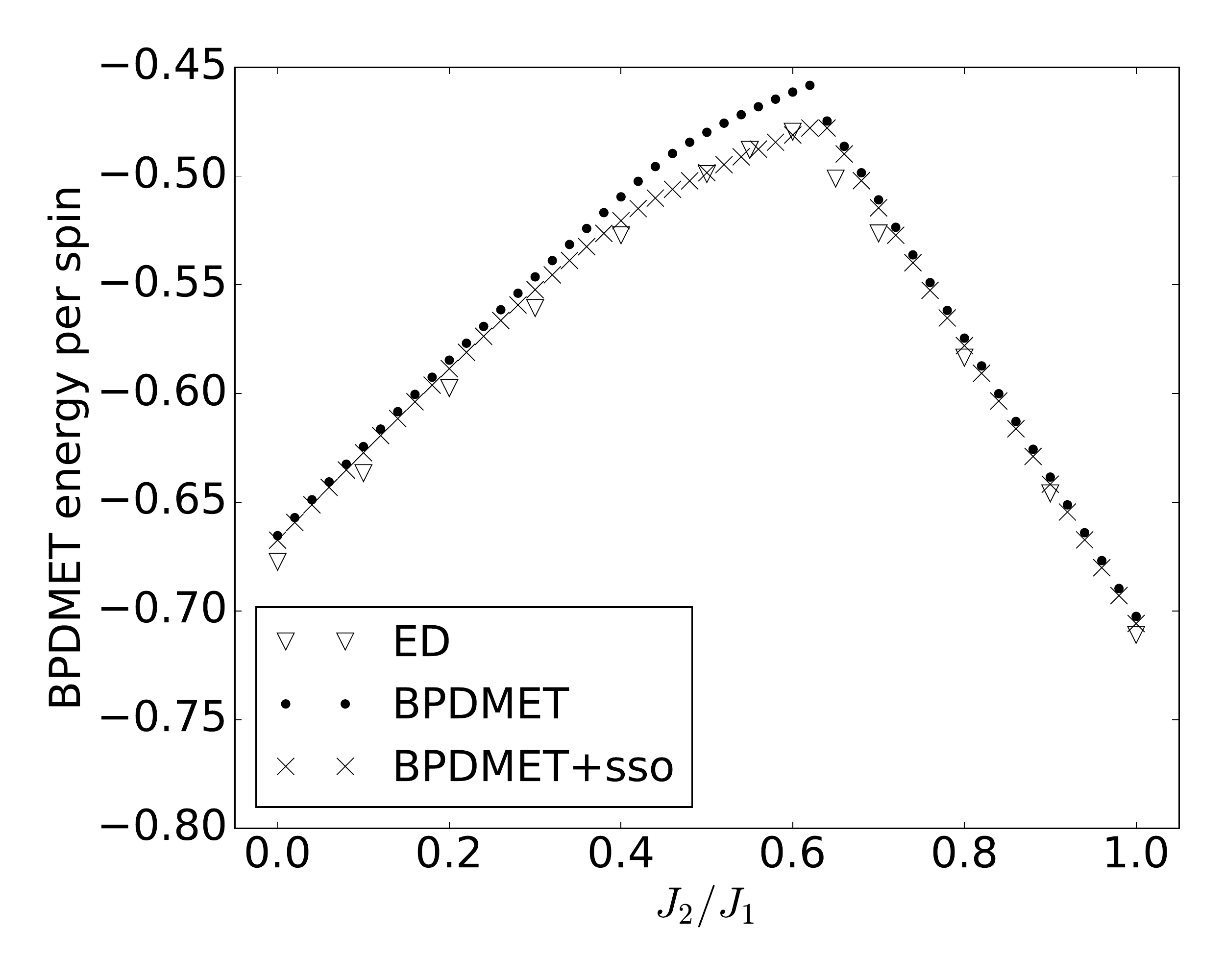}
\caption{\label{fig:sq_E40}BPDMET energy of the 40-spin square lattice with NN and NNN interaction using sweeps. Results both with and without spin-state optimization are given. Exact results are obtained from Ref. \onlinecite{Richter-2010}.}
\end{figure}

The BPDMET energies (see Eq.~(\ref{eq:DMETenergy})) are calculated for a $8\times8$ square lattice with periodic boundary conditions, i.e. an $8\times8$ square lattice on a torus. The lattice is divided into 16 equal $2\times2$ clusters of which one is chosen as impurity. Ground-state energies are calculated as explained in section \ref{sec:expval}. In Fig.~\ref{fig:sq_random}, the converged energy values for BPDMET with random initialization are given. At high and low $J_2/J_1$ convergence of the BPDMET algorithm happens quite consistently to the same energy values. When $J_2/J_1 \in [0.6,0.8]$, multiple energy values are found and the algorithm converges to a variety of local minima. For purposes of reproducibility, we will make use of sweeps (Fig.~\ref{fig:sq_64} and Fig.~\ref{fig:sq_E40}). A sweep starts in a region where convergence is consistent to the same minimum, and sweeps through the parameter region using previous converged results as initialization. Sweeps can be done from low to high $J_2/J_1$ or vice versa. The BPDMET energy is used as selection criterion for the optimal solution. It should be stressed that using the random initializations (40 runs per parameter value) we never found a lower energy than the minimum energies found through sweeps from the left or right. This suggests that the sweep finds more optimal solutions that are hard to find with random initialization. This poses some justification for the use of sweeps.

Introduction of spin-state optimization in BPDMET yields significant changes in the calculated energy, as can been seen in Fig. \ref{fig:sq_64}(a). In the two ordered phases, minor changes in energy per spin are observed. In the intermediate paramagnetic phase, larger changes are visible. It is good to note that the BPDMET energy is not variational so a lowering in energy is not necessarily a net improvement of the energy. However, a study has been done on the 40-spin lattice (Fig.~\ref{fig:sq_E40}) and compared with exact results obtained in Ref. \onlinecite{Richter-2010}. The boundary conditions are chosen in the same way as in Ref. \onlinecite{Richter-2010}. When introducing the spin-state optimization for the 40-spin system, there is a substantial improvement in energy observed in the intermediate paramagnetic phase. This makes us confident that the substantial change in BPDMET energy through introduction of spin-state optimization is also a net improvement for the 64-spin lattice. 
%For instance, in a study of 16 spins it is noted that the BPDMET energy with spin-state optimization is lower than the exact value at intermediate $J_2/J_1$.

When looking at the variational energy $\lambda$ (Fig.~\ref{fig:sq_64}(a)), only a small correction occurs with the introduction of spin-state optimization. At $J_2 = 0$, $\lambda$ changes from $-0.5939$ per spin to $-0.5940$ per spin, while the non-variational BPDMET energy $E$ changes from $-0.6657$ per spin to $-0.6678$ per spin. The change in the BPDMET energy is clearly much larger than the change in the variational energy, and this was found to occur for all $J_2/J_1$ values. The improvement of the results is largely contained in the impurity spins. The energy per spin at $J_2 = 0$ ($E = -0.667840$) corresponds reasonably well with results obtained by quantum Monte Carlo (QMC) for the $8\times8$ lattice ($E = -0.673487$).\cite{Sandvik-1997}

Fan \textit{et al.}\cite{Fan-2015} calculated the energy per spin at $J_2 = 0$ as $E = 2\ev{\hat{\mathbf{S}}_1\cdot\hat{\mathbf{S}}_2}{\psi}$, making use of rotational symmetry present in the N\'eel phase. This is twice the bond energy between two NN spins in the impurity cluster. Using this expression for the energy per spin, $E = -0.66917$ and $E = -0.67045$ without and with spin-state optimization are obtained. Correspondence with the QMC solution clearly improves using this alternative energy expression.

The N\'eel and the collinear order parameters are calculated in order to identify the different phase transitions. These order parameters are given by
\begin{equation}
M^2_N(\mathbf{Q}) = \frac{1}{N^2}\sum\limits_{ij} \langle \hat{\mathbf{S}}_i \cdot \hat{\mathbf{S}}_j \rangle e^{i \mathbf{Q}(\mathbf{R}_i - \mathbf{R}_j)},
\end{equation}
where $\mathbf{Q}$ is given by $(\pi,\pi)$ for the N\'eel parameter and by $(\pi, 0)$ and $(0,\pi)$ for the collinear parameter in $x$- and $y$-direction. $R_i$ is the position vector of the spin site $i$ and $N$ is the total number of spins. For an infinite lattice rotational symmetry breaking will occur. However, the exact solution for the finite lattice (for instance a $4\times4$ lattice) yields a collinear order parameter that is equal in the $x$- and $y$-direction. We notice that BPDMET finds a rotation symmetry broken solution, even in finite lattices.
In Fig. \ref{fig:sq_64}(b) the order parameters are shown. The largest collinear parameter is plotted here.

When introducing spin-state optimization, small changes are noticeable in the order parameters. The N\'eel order parameter at $J_2 = 0$ ($M^2_N = 0.19825$ and $M^2_N = 0.19670$ without and with spin-state optimization) corresponds well with results obtained through QMC ($M^2_N = 0.17784$).\cite{Sandvik-2010, Sandvik-1997} Phase transitions at $J_2/J_1 \approx 0.4$ and $J_2/J_1 \approx	0.62$ can be observed, in correspondence with previous studies. The location of these phase transitions does not change with the introduction of spin-state optimization. A strong N\'eel order is observed at low $J_2/J_1$ while a strong collinear order is observed at high $J_2/J_1$. In the intermediate region, both order parameters stay rather small, but do not completely vanish due to finite size effects.

\subsection{\label{sec:kitheis}The Kitaev-Heisenberg model}

\begin{figure}[ht]
\includegraphics[width = 0.48\textwidth]{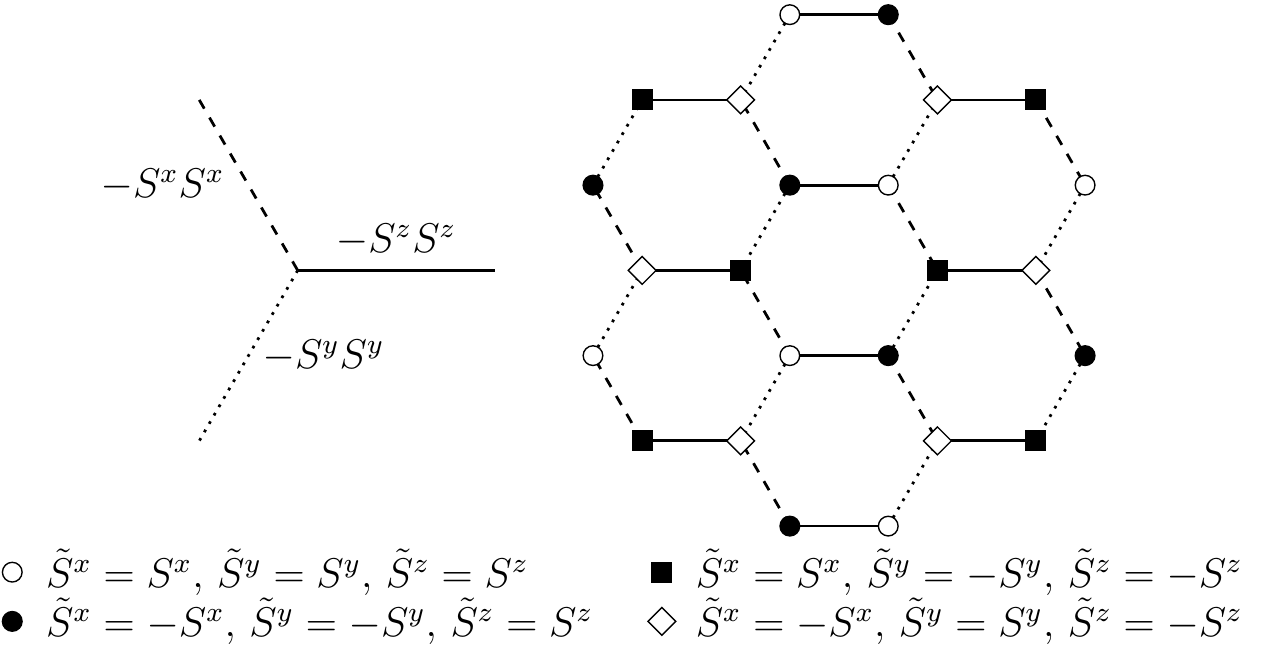}
\caption{$x$-, $y$- and $z$-links for the Kitaev terms on the honeycomb lattice with 24 spins. The dashed, dotted and full links are the $x$-, $y$- and $z$-links, respectively. The 4-sublattice rotated basis\cite{Chaloupka-2010} $\hat{\tilde{\mathbf{S}}}$ is also shown.}
\label{fig:honeylattice}
\end{figure}

The applicability of BPDMET is not limited to square lattices, other lattices are also equally feasible. In this section we consider the Kitaev-Heisenberg model\cite{Chaloupka-2010, Jiang-2011, Reuther-2011, Price-2012, Chaloupka-2013, Oitmaa-2015} on the honeycomb lattice. We study a 24-spin lattice with periodic boundary conditions (see Fig.~\ref{fig:honeylattice}). For this system exact diagonalization of the system is still feasible and the BPDMET method can be benchmarked. This model is a mixture of the Kitaev model\cite{Kitaev-2006} and the Heisenberg model on the honeycomb lattice and spin interactions are given by the Hamiltonian:
\begin{equation}
\begin{split}
\hat{H} = -J_1 &\sum\limits_{x-\mathrm{links}}\hat{S}^x_i \hat{S}^x_j -J_1 \sum\limits_{y-\mathrm{links}}\hat{S}^y_i \hat{S}^y_j\\
-J_1 &\sum\limits_{z-\mathrm{links}}\hat{S}^z_i \hat{S}^z_j
+J_2 \sum\limits_{\langle ij \rangle} \hat{\mathbf{S}}_i \cdot \hat{\mathbf{S}}_j.
\end{split}
\end{equation}

$J_1$ represents the strength of the anisotropic Kitaev interaction while $J_2$ is the strength of the Heisenberg part. Every spin is connected through exactly one $x$-, $y$- and $z$-link to a neighboring spin. The different links are shown in Fig. \ref{fig:honeylattice}.

The Kitaev and Heisenberg interaction are parametrized as $J_1 = 2\alpha$ and $J_2 = 1 - \alpha$. In the interval $\alpha \in [0,1]$, three phases occur.\cite{Chaloupka-2010} At low $\alpha$ a N\'eel AF phase is observed, at intermediate $\alpha$ a stripy AF phase, and at high $\alpha$ a quantum spin liquid is observed. The phase transitions are expected at $\alpha \approx 0.4$ and $\alpha \approx 0.8$.\cite{Chaloupka-2010, Jiang-2011, Reuther-2011, Price-2012, Chaloupka-2013, Oitmaa-2015} At the intermediate point, $\alpha = 0.5$, the system is exactly solvable through the use of a rotated basis.\cite{Chaloupka-2010} This rotated basis is defined by dividing the lattice into 4 sublattices and defining different rotated spin operators $\hat{\tilde{\mathbf{S}}}$ on these sublattices as can be seen in Fig.~\ref{fig:honeylattice}. At this intermediate point, the Hamiltonian is reduced to the ferromagnetic Heisenberg model in the rotated basis, and exact solutions are given by states with maximal total spin in the rotated basis (e.g. $\ket{\uparrow\uparrow\dots\uparrow}$). The two trivial states with maximal total spin can be clearly represented by the BPDMET ansatz. The basis transformation suggested in Ref. \onlinecite{Chaloupka-2010} leaves the corner of the Hilbert space described by the BPDMET ansatz unchanged.\cite{Gunst-2016} In the original basis it is therefore also possible to find exact solutions at this intermediate point through BPDMET. 

\begin{figure}
\includegraphics[width = 0.48\textwidth]{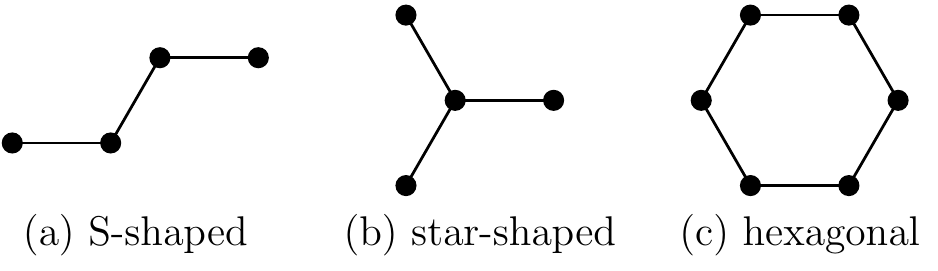}
\caption{Different possible cluster-coverings for the hexagonal lattice.}
\label{fig:honey_clusters}
\end{figure}

\begin{figure}
\includegraphics[width = 0.48\textwidth]{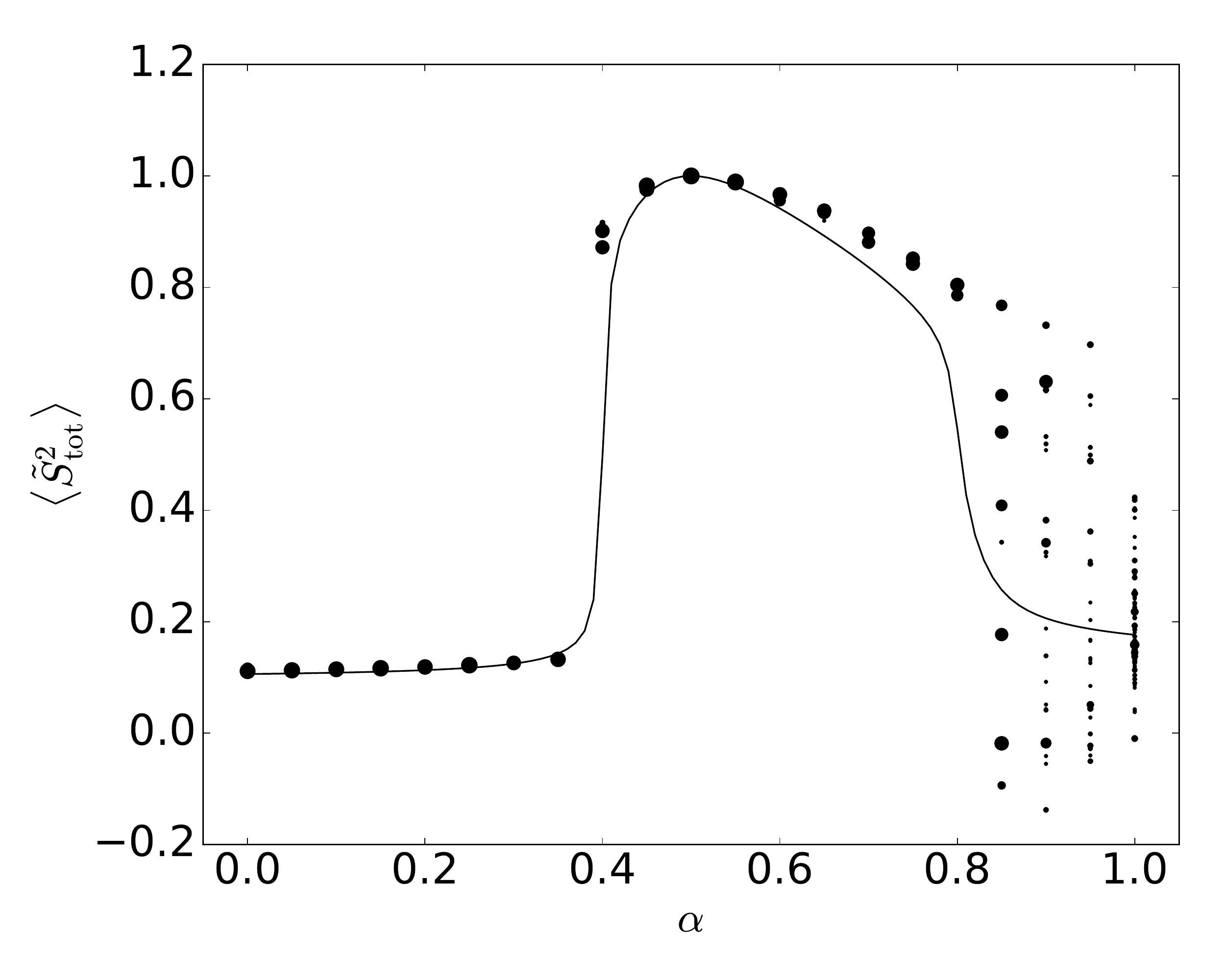}
\caption{Calculations executed with random initialization. Calculations are executed with the S-shaped cluster (Fig. \ref{fig:honey_clusters}(a)). At each $\alpha$ value, 40 random calculations are executed. The squared total spin in the rotated basis is given. The size of the markers scale with the number of calculations that yielded the corresponding value. Since the calculated squared total spin is not a squared total spin of a wave function but calculated as described in section \ref{sec:expval}, the values can be negative. The full line is the exact result for 24 spins.}
\label{fig:S_random}
\end{figure}

BPDMET allows freedom in the choice of the impurity and bath clusters. Three different types of clusters that can cover the entire lattice (see Fig.~\ref{fig:honey_clusters}) have been examined. 

When randomly initializing the BPDMET algorithm, convergence happens quite consistently in the two AF phases ($\alpha < 0.8$); however, in the spin liquid phase, the calculations converge to a wide variety of local minima as can be seen in Fig.~\ref{fig:S_random} with the calculation of $\hat{\tilde{\mathbf{S}}}^2_\mathrm{tot}$ for the S-shaped cluster. Similar results were obtained for the star shaped and hexagonal clusters. It was found that results obtained with the S-shaped cluster are slightly inferior to the other two types of clusters in describing the phase transitions. We will therefore only present these random initialization results for the S-shaped cluster and in the remainder of the paper only the star shaped and hexagonal clusters are discussed.

\begin{figure}
\includegraphics[width = 0.45\textwidth]{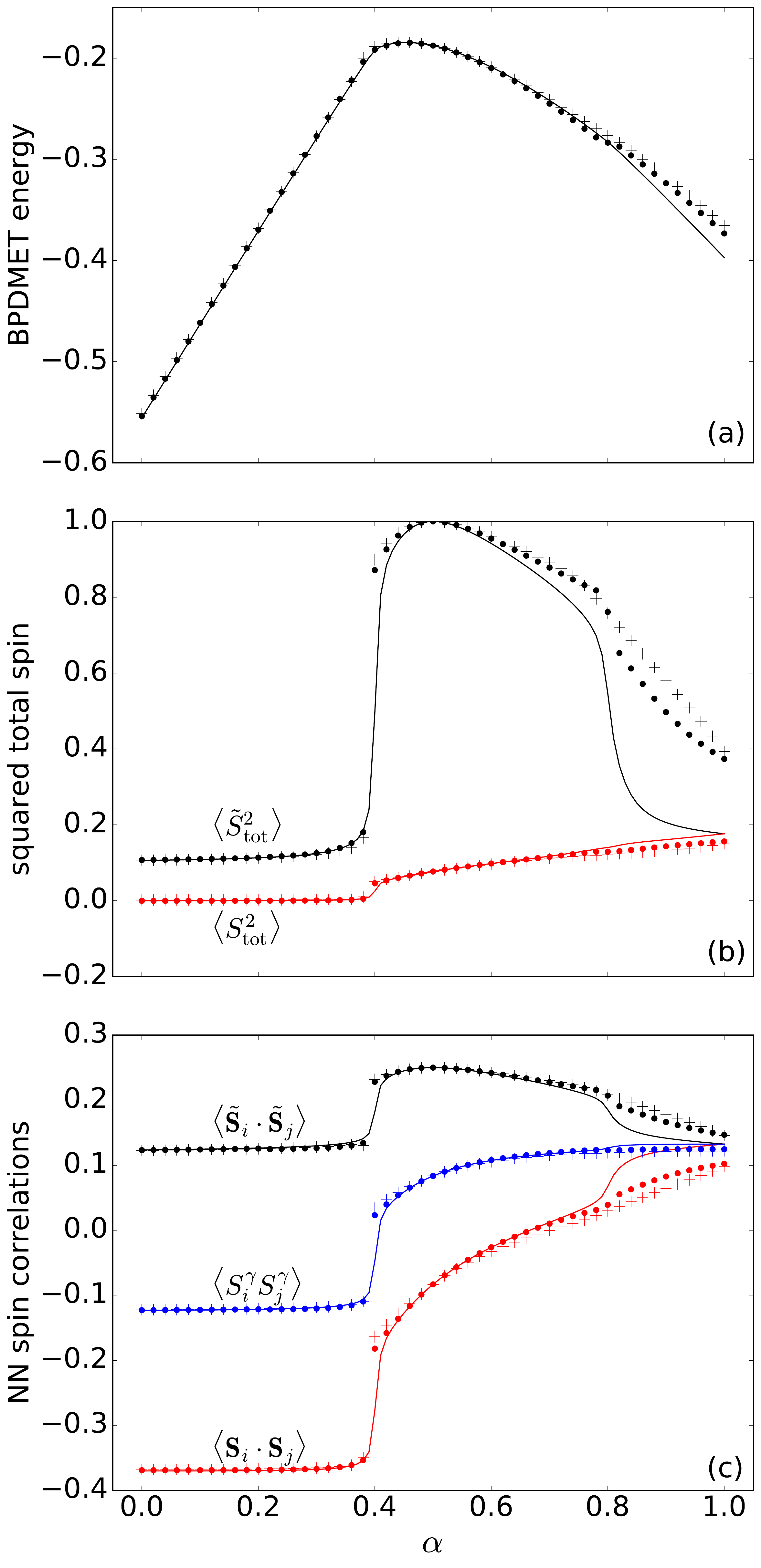}
\caption{(Color online) BPDMET results for the star-shaped cluster (Fig. \ref{fig:honey_clusters}(b)). Crosses are results without and dots are results with spin-state optimization. Full lines are exact results for the 24 spin lattice. (a) Energy per spin. (b) Squared total spin in original ($\langle S^2_\mathrm{tot} \rangle$) and rotated basis ($\langle \tilde{S}^2_\mathrm{tot} \rangle$). (c) Nearest neighbor spin spin correlations in original ($\langle \mathbf{S}_i \cdot \mathbf{S}_j \rangle$) and rotated basis ($\langle \tilde{\mathbf{S}}_i \cdot \tilde{\mathbf{S}}_j \rangle$) and nearest neighbor spin spin correlation in the bond direction ($\langle S_i^\gamma S_j^\gamma \rangle$).}
\label{fig:star_plots}
\end{figure}

\begin{figure}
\includegraphics[width = 0.45\textwidth]{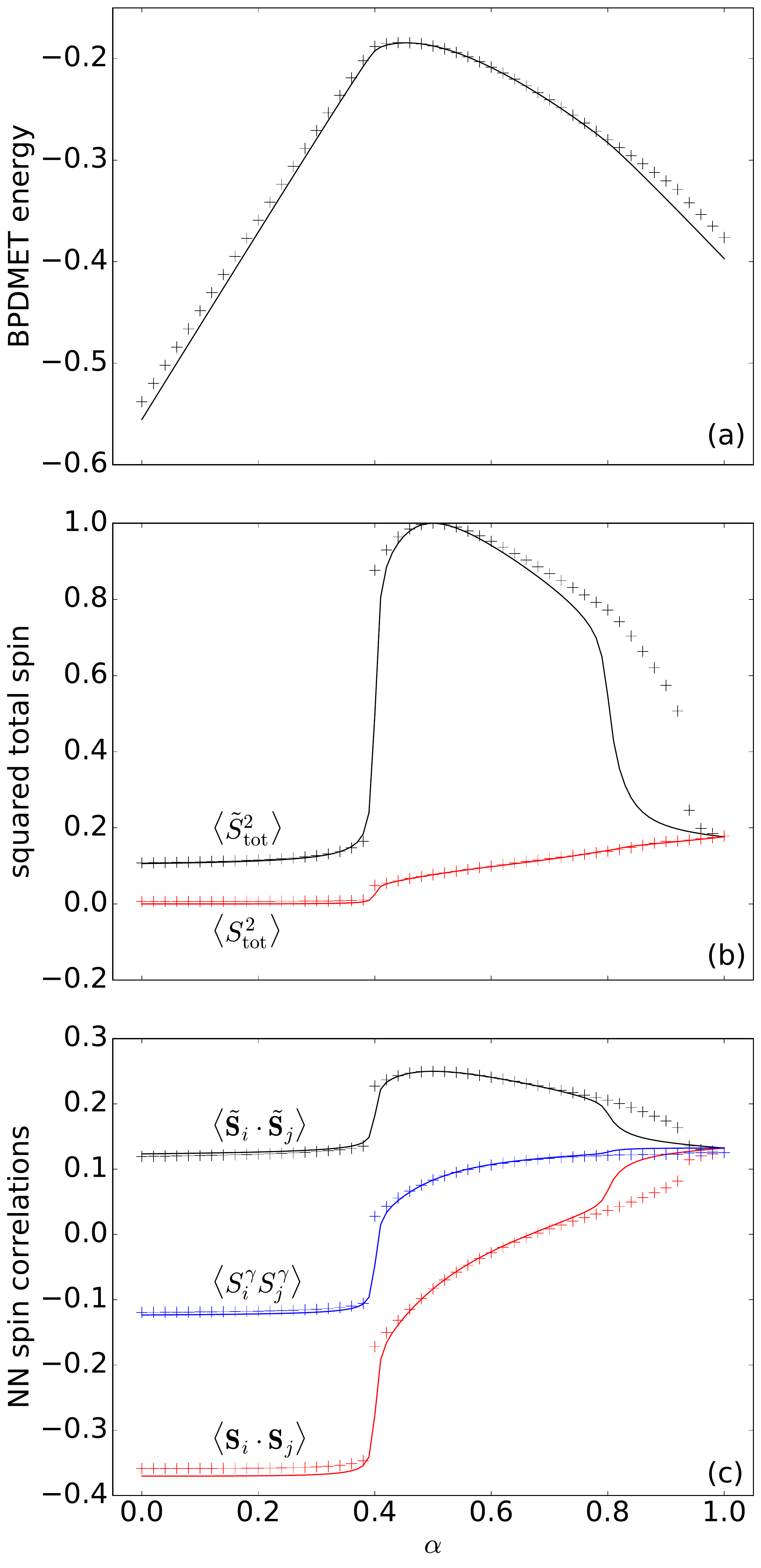}
\caption{(Color online) BPDMET results for the hexagonal cluster (Fig. \ref{fig:honey_clusters}(c)). Crosses are results without spin-state optimization. Full lines are exact results for the 24 spin lattice. (a) Energy per spin. (b) Squared total spin in original ($\langle S^2_\mathrm{tot} \rangle$) and rotated basis ($\langle \tilde{S}^2_\mathrm{tot} \rangle$). (c) Nearest neighbor spin spin correlations in original ($\langle \mathbf{S}_i \cdot \mathbf{S}_j \rangle$) and rotated basis ($\langle \tilde{\mathbf{S}}_i \cdot \tilde{\mathbf{S}}_j \rangle$) and nearest neighbor spin spin correlation in the bond direction ($\langle S_i^\gamma S_j^\gamma \rangle$).}
\label{fig:hex_plots}
\end{figure}

Sweeps, either from the left or from the right, are again used to ensure reproducibility of the results.

Comparison of the BPDMET results for the 24-spin system with the exact results shows a difference between the star-shaped and hexagonal clusters. Both clusters describe the two AF phases well.  When using the star-shaped cluster a phase transition occurs at $\alpha \approx 0.8$ (Fig. \ref{fig:star_plots}). Although the phase transition to the spin liquid is observed, the spin liquid itself is poorly described through BPDMET as can be inferred from the calculated BPDMET energies (Fig.~\ref{fig:star_plots}(a)), spin properties (Fig.~\ref{fig:star_plots}(b)) and correlation functions (Fig.~\ref{fig:star_plots}(c)). The phase transition to the spin liquid gets more pronounced when introducing spin-state optimization, but the spin liquid itself is still not represented adequately. 

In general, the BPDMET ansatz can capture a larger corner of the Hilbert space with increasing impurity size, so it is expected that the hexagonal cluster performs better than the 4-spin clusters. When using the hexagonal cluster, the phase transition towards a spin liquid is detected and the obtained properties of the spin liquid are in good correspondence with the exact diagonalization at $\alpha = 1$ (Fig. \ref{fig:hex_plots}).
However, the phase transition happens at $\alpha \approx 0.92$, which is not the right value. Calculations with spin-state optimization have not been performed for this type of cluster since they were already quite intensive without the optimization.

BPDMET allows to investigate larger systems than the 24 spin lattice. When expanding the honeycomb lattice by one extra layer, a 54 spin honeycomb lattice is obtained which is however not coverable with the S-shaped and star shaped clusters. Also, a rotated basis respecting periodic boundary conditions as proposed in Ref. \onlinecite{Chaloupka-2010}, cannot be found. With another extra layer, we get a 96 spin honeycomb lattice that is coverable with the three types of clusters (Fig.~\ref{fig:honey_clusters}), and is also consistent with the concept of the rotated basis. No shift in the location of the phases is found when extending to 96 spins.

\subsection{Spectral functions}

In section \ref{sec:tangent}, a method is introduced for the BPDMET to find the excitations and the spectral functions through the tangent space. For the calculation of the spectral function Eq.~(\ref{eq:spectral}) is used, with $\eta = 0.01$. In Eq.~(\ref{eq:spectral}), two possible energy values for $E_n$ can be used, i.e. the variational energy $\lambda$ and the BPDMET energy as given in Eq.~(\ref{eq:DMETenergy}).

For the Kitaev-Heisenberg model at $\alpha = 0.5$, the exact solution can be found through BPDMET. We expect thus that the spectral function will be reproduced quite well. Since the ground state in $\alpha =0.5$ is degenerate, the equation for the spectral function is changed to
\begin{equation}
\begin{split}
A(\omega, \hat{X}) = 
-\frac{1}{\pi d}\Im\left[\sum\limits_{i = gs}\sum\limits_{j \neq gs} \frac{|\mel{\phi_j}{\hat{X}}{\phi_i}|^2}{\omega - \Delta E_{j} + i\eta}\right],
\end{split}
\label{eq:degen_spec}
\end{equation}
where $d$ is the dimension of the degenerate ground state space, $\Delta E_{j}$ is the energy difference between the state $\ket{\phi_j}$ and the energy of the degenerate ground states, and the index $i$ sums over all ground states while $j$ sums over the excited states. This expression is independent of the chosen basis in the degenerate ground state space.
\begin{figure}
\includegraphics[width = 0.48\textwidth]{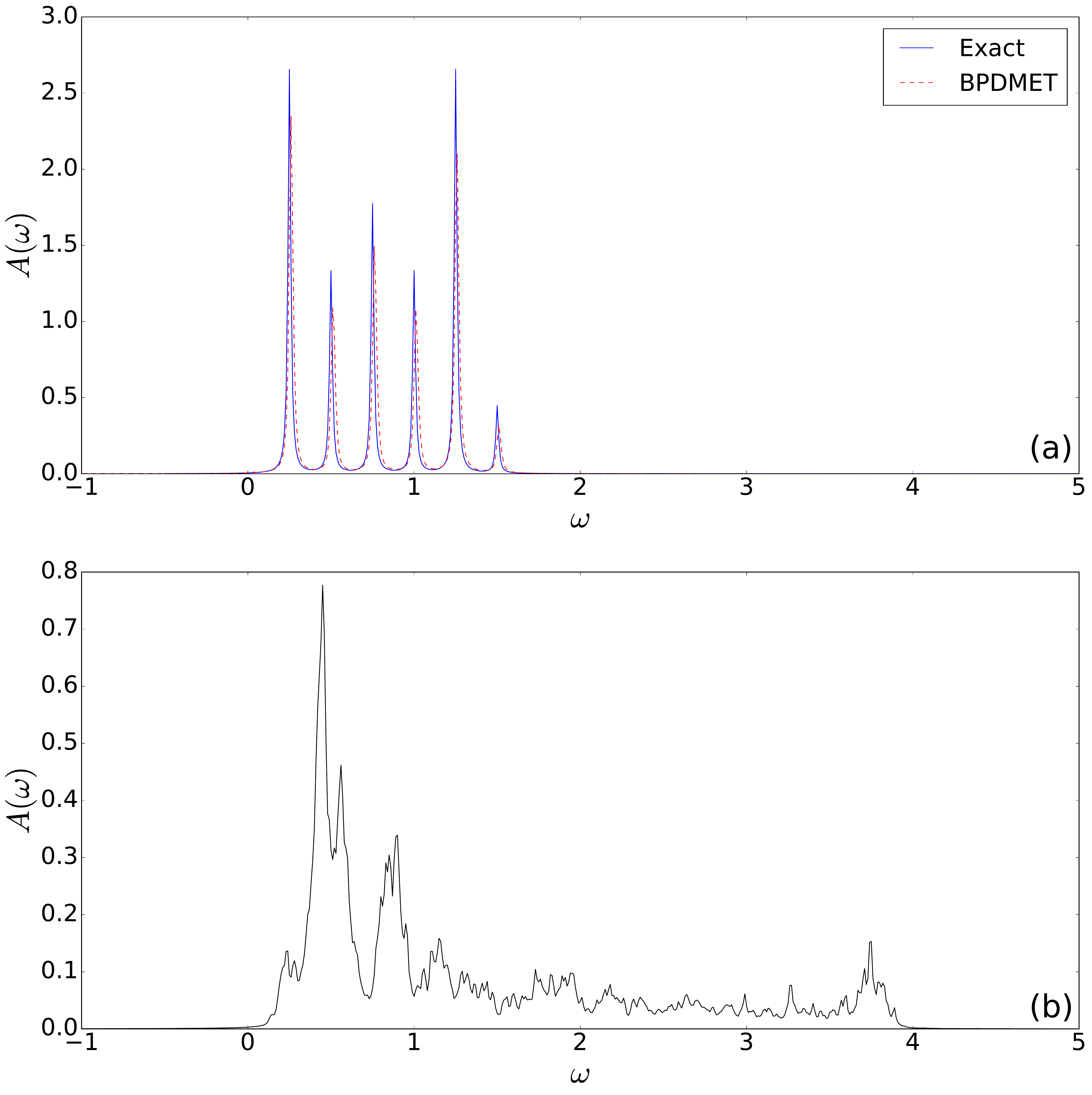}
\caption{(Color online) The spectral function for the Kitaev-Heisenberg model with 24 spins at $\alpha = 0.5$. The perturbation $\hat{X}$ is chosen as $\hat{S}^+$ and $\eta = 0.01$. The BPDMET spectral function is averaged out over 100 calculations with random initialization. (a) shows the exact results (full line) and results obtained through BPDMET with spin-state optimization where the variational energy $\lambda$ is used for the $E_n$ values (dotted line). (b) shows the results obtained through BPDMET with spin-state optimization where the non-variational BPDMET energy $E$ is used for the $E_n$ values.}

\label{fig:exact_spectr}
\end{figure}

\begin{figure}
\includegraphics[width = 0.48\textwidth]{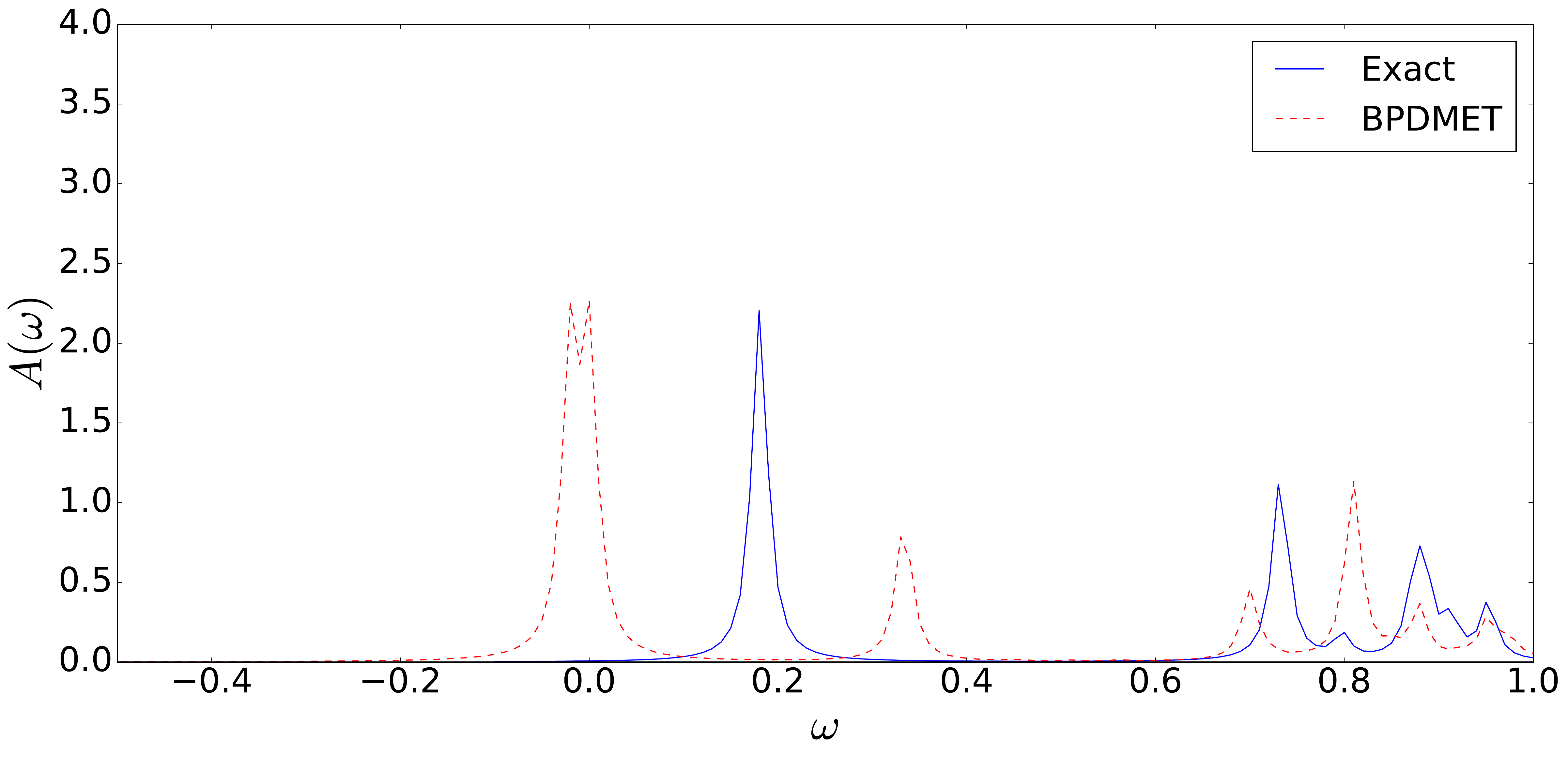}
\caption{(Color online) The spectral function for the Kitaev-Heisenberg model with 24 spins at $\alpha = 0.25$. The perturbation $\hat{X}$ is chosen as $\hat{S}^z$ and $\eta = 0.01$. Exact results (full line) are compared with results obtained through BPDMET with spin-state optimization (dotted line).}
\label{fig:approx_spectr}
\end{figure}

To calculate the spectral function through BPDMET we simply perform the calculations multiple times with random initializations and average out over the obtained spectral functions. Residues between the ground states are not taken into account in the spectral function, as can be seen in Eq.~(\ref{eq:degen_spec}). To make sure that the tangent space method does not take these residues into account, only eigenvectors with an eigenvalue significantly different from the ground state energy are used in the summation in Eq.~(\ref{eq:spectral}). When diagonalizing the Hamiltonian in the tangent space, a few eigenvalues very close to the ground state energy are obtained (within a 0.03 margin), which are separated from the other excitations by a clear gap (the next eigenvalues are found at $\approx 0.25$). These eigenstates are left out of the spectral function since they originate from the degenerate ground-state solutions within the tangent space.

We calculate the spectral function with $\hat{S}^+_i$ used as perturbation $\hat{X}$, i.e. the raising operator on spin $i$. In Fig.~\ref{fig:exact_spectr}(a), it is observed that the tangent space method finds practically the exact spectrum when using the variational energy. Some small differences in peak location and amplitude are visible.

When using the non-variational DMET energy in Eq.~(\ref{eq:spectral}), all correspondence with the exact solution is lost (see Fig.~\ref{fig:exact_spectr}(b)). It is therefore clear that the variational energy should be used in the calculation of the spectral function in contrary to the calculation of ground state properties. There, the non-variational BPDMET energy has proven to be superior.

In the previous example, BPDMET is able to find the exact ground-state wave function. When this is not the case, the calculation of the spectral function through tangent space diagonalization can be inadequate. This can be seen in Fig. \ref{fig:approx_spectr}, where the spectral function for the same Kitaev-Heisenberg model is calculated at $\alpha = 0.25$. BPDMET is, in this case, unable to find the exact ground state. For this system, correspondence of the BPDMET spectral function with exact results is lost.

\section{\label{sec:Conclusions}Conclusion}

The BPDMET method is used to study the spin-1/2 anti ferromagnetic Heisenberg model on the square lattice with 64 spins as originally done in Ref. \onlinecite{Fan-2015} and the spin-1/2 Kitaev Heisenberg model\cite{Chaloupka-2010} on the honeycomb lattice with different sizes. A systematic approach for the calculation of properties within the BPDMET framework is introduced. Spin-state superposition in the impurity has been added to BPDMET yielding improved results for the energy profile and properties. The calculation of excited states and the spectrum through diagonalization in the tangent space of the BPDMET ground state is investigated, but provides somewhat unsatisfactory results. In any case, it has been shown that for the calculation of the spectral function the variational energy $\lambda$ should be used and not the BPDMET energy.

For the AF Heisenberg model on the square lattice, order parameters are calculated and the right phases are detected in the system. For the Kitaev Heisenberg model, different types of clusters are used. The results for a 24-spin honeycomb lattice are compared with the exact solution. It is clear that the results are dependent on the cluster shape and that some clusters are more appropriate for certain phases than others. However, none of the cluster shapes are able to represent the spin liquid regime. Only with hexagonal clusters, the spin liquid phase is detected, but the phase transition happens at a wrong value of the coupling parameters. BPDMET enables one to investigate larger systems unreachable with exact diagonalization. The ground state for the 96-spin honeycomb lattice is calculated and it is found that the position of the phase transitions does not change when enlarging the system.

\begin{acknowledgments}
K.G. and S.D.B. acknowledge support from the Research Foundation Flanders (FWO Vlaanderen).
\end{acknowledgments}

\appendix*
\section{\label{app:error}Convergence}

For a variational optimization, we show that the error on the energy scales quadratically with the error on the wave function. For expectation values of general properties, the error scales linearly with the wave function error. In order to see this, we decompose our approximate wave function into a component parallel to the exact ground state and a perpendicular error-component, i.e.\cite{Wouters-2014} 
\begin{equation}
\ket{\Psi} = \sqrt{1 - \epsilon ^ 2} \ket{\Psi_0} + \epsilon \ket{\Psi_\mathrm{error}}
\end{equation}
with $\braket{\Psi_0}{\Psi_\mathrm{error}} = 0$ and $\epsilon$ a measure for the error. The error on the wave function is
\begin{equation}
\begin{split}
|\ket{\Psi} - \ket{\Psi_0}||_2 &= \sqrt{\left(\sqrt{1 - \epsilon^2} - 1\right)^2 + \epsilon ^ 2 }\\
&= \epsilon + \mathcal{O}(\epsilon ^ 2)
\end{split}
\end{equation}
and thus linear in $\epsilon$ for small errors. We get 
\begin{equation}
\begin{split}
\ev{\hat{Q}}{\Psi} - \ev{\hat{Q}}{\Psi_0} =\\ \left(\mel{\Psi_\mathrm{error}}{\hat{Q}}{\Psi_0} + \mel{\Psi_0}{\hat{Q}}{\Psi_\mathrm{error}}\right) \epsilon \\
+ \left(\ev{\hat{Q}}{\Psi_\mathrm{error}} - \ev{\hat{Q}}{\Psi_0}\right) \epsilon^2 + \mathcal{O}(\epsilon^3)
\end{split}
\label{eq:error_prop}
\end{equation}
for the error on the expectation value of general properties. We see that this error is linear with $\epsilon$. Since $\ket{\Psi_0}$ is an eigenstate of $\hat{H}$ and $\braket{\Psi_\mathrm{error}}{\Psi_0} = 0$, the linear term vanishes for the error on the energy and the leading error term is thus quadratic in $\epsilon$.
In BPDMET, an optimization is performed in a restricted Hilbert space during each minor iteration step. During this optimization, the optimal solution within the restricted space is chosen by diagonalizing the effective Hamiltonian in this restricted Hilbert space (as discussed in section \ref{sec:opt}). The updated wave function is therefore an eigenstate (corresponding to the minimal eigenvalue) of the effective Hamiltonian (not necessarily of the full Hamiltonian).

In each minor iteration, the wave function $\ket{\Psi}$ is updated to the ground state $\ket{\Psi_0}$ of the effective Hamiltonian. The change during each minor iteration of the variational energy $\lambda$ is given by Eq.~(\ref{eq:error_prop}) with $\hat{Q} = \hat{H}_\mathrm{eff}$ and for the same reason as explained above, will have a quadratic leading term. The effective non-variational BPDMET energy operator (i.e. the BPDMET energy operator mapped to the restricted Hilbert space), will have a linear leading term.

When the BPDMET algorithm is close to a local minimum and close to convergence, the correction parameter $\epsilon$ will be small. The corrections on the variational energy $\lambda$ will be quadratic in $\epsilon$, while the corrections on the BPDMET energy will be linear in $\epsilon$. Using the BPDMET energy as convergence criterion is therefore more stringent.

\bibliography{mybib.bib}

\end{document}